\documentclass[conference,a4paper]{APSIPA2021}
\usepackage{multirow}
\usepackage[pdftex]{graphicx}
\usepackage{amsmath}
\usepackage[psamsfonts]{amssymb}
\usepackage{amsxtra}
\usepackage{threeparttable}
\usepackage{epstopdf}
\pdfoutput=1

\begin{document}

\title{TFCN: Temporal-Frequential Convolutional Network for Single-Channel Speech Enhancement}

\author{%
\authorblockN{%
Xupeng Jia and
Dongmei Li
}
\authorblockA{%
Tsinghua University, Beijing, China \\
E-mail: jxp12@mails.tsinghua.edu.cn, lidmei@tsinghua.edu.cn \\
Tel: +86-10-62782693}
}

\maketitle
\thispagestyle{empty}

\begin{abstract}
  Deep learning based single-channel speech enhancement tries to train a neural network model for the prediction of clean speech signal. There are a variety of popular network structures for single-channel speech enhancement, such as TCNN, UNet, WaveNet, etc. However, these structures usually contain millions of parameters, which is an obstacle for mobile applications. In this work, we proposed a light weight neural network for speech enhancement named TFCN. It is a temporal-frequential convolutional network constructed of dilated convolutions and depth-separable convolutions. We evaluate the performance of TFCN in terms of Short-Time Objective Intelligibility (STOI), perceptual evaluation of speech quality (PESQ) and a series of composite metrics named Csig, Cbak and Covl. Experimental results show that compared with TCN and several other state-of-the-art algorithms, the proposed structure achieves a comparable performance with only 93,000 parameters. Further improvement can be achieved at the cost of more parameters, by introducing dense connections and depth-separable convolutions with normal ones. Experiments also show that the proposed structure can work well both in causal and non-causal situations.
\end{abstract}

\section{Introduction}
Single-channel speech enhancement tries to suppress the background noise while keeping the distortion of the target speech as low as possible. It usually aims at improving the perceptual quality and intelligibility of speech, or reducing the word error rate of a speech recognition system. Speech enhancement algorithms are widely used in speech-related applications and have been studied for several decades. There are numerous reliable speech enhancement approaches based on traditional signal processing, and they have been already used in our daily life \cite{wiener, spectral_subtraction, mmse}. However, it’s still a great challenge for them to deal with the unstable noise and complex situations.

With the development of deep learning techniques, speech enhancement algorithms based on deep neural networks (DNNs) have shown their advantages over the classical ones. A number of DNN-based speech enhancement algorithms have been proposed in the literature. They try either to predict masks such as IBM \cite{IBM}, IRM \cite{IRM} or PSM \cite{PSM}, or to learn a mapping function from the noisy input to the target signal \cite{spec_mapping, PL_learning, time-domain}. The network structure is an important factor that affects the performance of a speech enhancement algorithm. A variety of structures have been applied in the speech enhancement task, such as Fully connected networks \cite{DNN}, Convolutional Neural Networks (CNNs) \cite{CNN}, Recurrent Neural Networks (RNNs) \cite{RNN} and their combinations \cite{CRN}.

TCN was proposed in the action augmentation task at first \cite{TCN}, and has been shown to be an effective structure for speech enhancement and separation. Luo proposed ConvTasNet \cite{convtasnet}, which employs a TCN as the major functional block and surpasses ideal TF magnitude masking in speech separation performance. Pandey proposed TCNN, which has an encoder-decoder structure similar to U-Net, and utilizes a TCN block between the encoder and the decoder \cite{tcnn}. TCNN was shown to have a better performance than the convolutional recurrent neural network (CRN) \cite{CRN}. Kishore applied the ConvTasNet in speech enhancement, and proposed a multi-layer encoder-decoder structure for TCN \cite{multilayerEncoder}. Besides the TCN-based speech enhancement algorithms in time-domain, Zhang proposed a TCN-based method in TF-domain named Multi-scale TCN \cite{multiscale}. It uses the TCN to learn a mapping function of Log Power Spectrum (LPS) between noisy and clean speech signals, and adopts skip connections from the noisy input to each residual block in TCN. The performance advantage of TCN comes from its large and flexible receptive filed and capability to learn the local and long-time context information. Besides, the structure of TCN usually has a smaller model size compared with other structures, due to the application of point-wise and depth-wise convolutions.

When applying TCN directly in TF-domain speech enhancement, it results in a convolutional operation along the time axis and a fully connected operation along the frequency axis. As described in \cite{CRN}, fully connected networks have difficulties in tracking the target speaker in speech separation. Besides, as the number of convolution channels are usually set to several hundred to guarantee a satisfactory performance, the fully connected operations introduce a large number of parameters and are very unefficient. UNet \cite{unet} is another popular network structure in speech enhancement, applying convolutional operations along both the time and frequency axes. Recent studies show that LPS mapping system using UNet and its improved version \cite{affinity} achieves state-of-the-art performance. However, their performances may be limited by the small receptive fields, and the model sizes of them are very large. In this work, we propose TFCN by simply replacing the 1-D convolutions in TCN with 2-D convolutions. Except for the application of point-wise and depth-wise convolutions, we limit the number of feature maps to further reduce the model size. We employ dense connections to improve the performance and evaluate the algorithm under causal, non-causal and semi-causal situations.

The rest of the paper is organized as follows. Structures of TCN and the proposed TFCN are described in section 2. Experimental setups and results are presented in section 3 and conclusions are drawn in section 4.

\section{Proposed system}

\subsection{System overview}

\begin{figure}[t]
\begin{center}
\includegraphics[width=85mm]{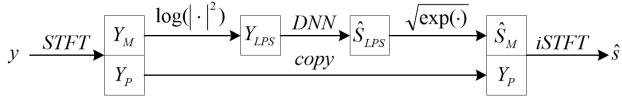}
%\includegraphics[width=70mm]{Fig1-TimesNewRomanPSMT.eps}
%\fbox{ \rule[-17mm]{0pt}{30mm} FIGURE }
\end{center}
\caption{System overview.}
% \vspace*{-3pt}
\label{fig:system_overview}
\end{figure}

% \begin{figure}[t]
%   \centering
%   \includegraphics[width=\linewidth]{fig_1_system_diagram.png}
%   \caption{System overview.}
%   \label{fig:system_overview}
% \end{figure}

The problem of single-channel speech enhancement can be formulated as
\begin{equation}
  y = s + n
  \label{eq1}
\end{equation}
where $y$, $s$ and $n$ are the observed noisy speech, the unknown target speech and the unknown noise, respectively. The aim of the single-channel speech enhancement system is to estimate $s$ from $y$, as precisely as possible. The system diagram is shown in Figure~\ref{fig:system_overview}. $y$ is the input signal and $\hat{s}$ is the estimated target signal. The subscript $M$ donates magnitude, $P$ donates phase and $LPS$ donates log power spectrum. The DNN is the core of the system, mapping the LPS of the noisy signal $Y_{LPS}$ into the LPS of the estimated signal $\hat{S}_{LPS}$, which is then combined with the noisy phase $Y_p$ to generate the estimated result $\hat{s}$. Note that although this system structure has been used in previous works, the different design of DNN results in different performance.

\subsection{Structure of TCN}

\begin{figure}[t]
\begin{center}
\includegraphics[width=85mm]{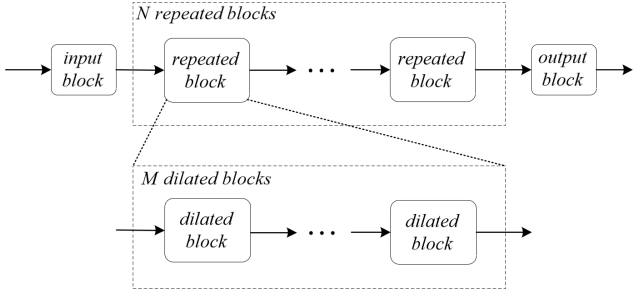}
%\includegraphics[width=70mm]{Fig1-TimesNewRomanPSMT.eps}
%\fbox{ \rule[-17mm]{0pt}{30mm} FIGURE }
\end{center}
\caption{Diagram of the TCN used in this work.}
% \vspace*{-3pt}
\label{fig:TCN_diagram}
\end{figure}
% \begin{figure}[t]
%   \centering
%   \includegraphics[width=\linewidth]{fig_2_TCN_diagram.png}
%   \caption{Diagram of the TCN used in this work.}
%   \label{fig:TCN_diagram}
% \end{figure}
%
\begin{figure}[t]
\begin{center}
\includegraphics[width=85mm]{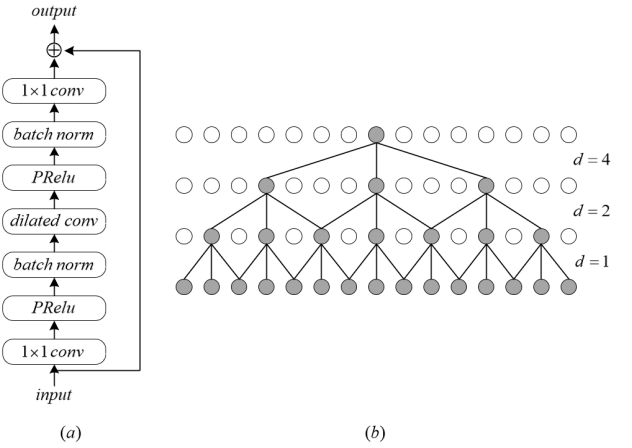}
%\includegraphics[width=70mm]{Fig1-TimesNewRomanPSMT.eps}
%\fbox{ \rule[-17mm]{0pt}{30mm} FIGURE }
\end{center}
\caption{Details of the dilated blocks. (a) Structure of the dilated block. (b) An example of dilated convolution, d donates dialtion rate.}
% \vspace*{-3pt}
\label{fig:dilated_block}
\end{figure}

% \begin{figure}[t]
%   \centering
%   \includegraphics[width=\linewidth]{fig_3_dilated_block.png}
%   \caption{Details of the dilated blocks. (a) Structure of the dilated block. (b) An example of dilated convolution, d donates dialtion rate.}
%   \label{fig:dilated_block}
% \end{figure}

TCN is a fully convolutional network structure that was proposed in 2016. It has shown its advantages on long sequence modeling and has also been successfully applied in speech separation and enhancement. The main idea of TCN is to perform convolutional operations along the time axis to utilize the context information. To further improve the length of the context window, long receptive field is achieved by operations such as max-pooling and dilated convolutions.

There are different versions of TCN realization. Figure~\ref{fig:TCN_diagram} is the diagram of the TCN used in this work, which is similar to the one used in TCNN \cite{tcnn}. The TCN is constructed of three modules, which are the input module, the enhancement module, and the output module. The input module is a batch normalization layer followed by a 1-D convolutional layer to change the number of feature maps. The output module is a 1x1 convolutional layer followed by PReLU non-linearity. The enhancement module is a series of repeated blocks that have the same structure. Each repeated block contains $M$ dilated blocks. The details of the dilated blocks are shown in Figure~\ref{fig:dilated_block}(a).

The dilated block includes three convolutional layers, which are the input 1x1 convolutional layer, the dilated convolutional layer and the output 1x1 convolutional layer. The 1x1 convolutional layer here is actually a 1-D point-wise convolutional layer, whose kernel size is $1$. PReLU non-linearity and batch normalization are applied after the input and dilated convolutional layers. The input convolutional layer and the output convolutional layer are used to change the number of the feature maps, while the dilated convolutional layer uses the dilated convolution to extract the context information. An example of the dilated convolution is shown in Figure~\ref{fig:dilated_block}(b). Assuming that the kernel size of the general convolutional layer is $3$, and the dilation rate is $d$, it means that $d-1$ zeros are inserted between any two adjacent elements and results in a dilated kernel size of $2d+1$. In TCN, the dilation rate of the dilated blocks in a repeated block are often set to a geometric progression, i.e. $1, 2, 4, 8, …$ Note that although the dilated convolutional layers are often realized by depth-wise convolution to reduce the number of parameters, there is a trade off between the model size and the performance. TCN with general convolutional layers for the dilated blocks achieves better performance than the one using depth-wise convolutional layers at the cost of more parameters.

\subsection{Structure of TFCN}

Although TCN can be used to TF-domain speech enhancement directly by treating the spectrogram as a time sequence and the frequency bins as different channels \cite{multiscale}, the fully connected manner along the frequency axis may limit its performance. 2-D convolution has already been used in many speech enhancement algorithms. It can utilize the temporal and frequential information at the same time, and effectively recover the TF patterns in spectrogram, even with a relatively small context window, which has been proved in previous works \cite{unet, affinity}. With this consideration, we proposed TFCN combining the advantages of the long and flexible receptive field of TCN and the strong capability of TF modeling of UNet.

The structure of TFCN is the same with the one of the TCN that is shown in Figure~\ref{fig:TCN_diagram}, with the details of each block modified. The input block is also a batch normalization layer followed by a convolutional layer. Different from the input block of TCN, the convolutional layer here is a 2-D convolutional layer. Because the input of TFCN has only one channel, i.e. the LPS, and the convolutional layer here is designed to expand the number of channels, point-wise convolution is not suitable for this block. The width and height of the kernel are set to values that are larger than one, for example, 5 and 7. The output module is constructed with a 2-D point-wise convolution and PReLU non-linearity.

The enhancement module also follows the structure shown in Figure~\ref{fig:TCN_diagram} and Figure~\ref{fig:dilated_block}, with all the convolutional layers using 2-D convolution instead of 1-D convolution. The 2-D dilated convolution expands its kernel along both the time and frequency axes. The dilation rates for the width and height of the kernel can be set to different values respectively. However, it is necessary to keep a relatively large receptive field along the frequential axis within a single repeated block to extract the frequential information effectively. Note that it is also possible to construct the TFCN with 1-D convolution by applying two convolutional layers for the temporal and frequential convolutions respectively. However, it would cost more memory space for training and the model size would be much larger than TFCN (because it's larger than TCN).

To further improve the performance of TFCN, dense connections are also employed. There are two kinds of dense connections, i.e. inter- and intra- the repeated blocks. The inter- ones connect the output of a repeated block to the input of every later repeated block. The intra- ones have the same manner but are applied between the dilated blocks in the same repeated block. Note that even with the dense connections between the dilated blocks, the residual connection is still indispensable.

\subsection{Causal and semi-causal realization of TFCN}

Many speech-related applications concern about the latency of the algorithm. The TFCN described above is a non-causal algorithm and has a symmetric receptive field, which means it receives information with equal length of time from the past and the future. However, because the TFCN is a fully convolutional neural network, it is easy to get a causal or semi-causal realization by padding and clipping operation. Take the dilated convolution in figure 3(b) as an example. Considering the second row from the bottom, adding one point of zero at the left end and cutting off the point at the right end can result in a causal realization. This requires that when implementing the first convolutional layer, a padding of 2 zero elements at each sides of the input should be used and the output points related with the padding at the right end should be cut off. Same operations can be done on the following layers, with an adaptation of the padding and clipping length. To get a semi-causal implementation, clipping length at the right end should be determined by the intended look-ahead length, and clipping at the left end should guarantee that the input and output have the same length.

Note that this padding and clipping operation should only be applied along the temporal axis, with the parameters related to the frequential axis remaining unchanged. What’s more, the convolutional layer in the input block should also be considered, which is different from the TCN. The batch normalization layers have no impact on the causal property because they work with fixed parameters in the inference stage.

\section{Experiments}

\subsection{Datasets}

The popular VCTK dataset constructed by Valentini et al \cite{VCTK_Valentini} is used to evaluate the performance of TFCN. The dataset is publicly available at https://datashare.ed.ac.uk/handle/10283/2791. It contains clean and noisy speech signals sampled at 48 KHz. The signals are resampled to 16 KHz in our experiments.

The train data is constructed with clean speech signals of 28 speakers from the Voice Bank Corpus \cite{voice_bank}, including 14 male and 14 female speakers. There are about 400 utterances for each speaker. Ten types of noise are used to generate the train data, including two artificial ones which are babble and speech-shaped noise, and eight real noise recordings from the Demand database \cite{Demand}. Four signal-to-noise ratio (SNR) levels are used, which are 0 dB, 5 dB, 10 dB and 15 dB, resulting in 40 different noise conditions. In our work, the train data is randomly divided into two parts, one contains 10077 utterances for training and the other one contains 1495 utterances for validation.

For the test data, 2 speakers from the Voice Bank Corpus and 5 types of noise from the Demand database are selected. Note that all the speakers and noises used for test data are different from the ones for train data. The SNR levels are set to 2.5 dB, 7.5 dB, 12.5 dB and 17.5 dB.

\subsection{Objective metrics}

Five popular objective metrics are used to evaluate the performance of the proposed method. They are STOI \cite{stoi}, PESQ \cite{pesq}, Csig, Cbak and Covl \cite{composite}. STOI measures the intelligibility of the enhanced signal, and its value is between 0 and 1. PESQ is a broadly used metric for the perceptron quality of speech, and it provides a value in the range from -0.5 to 4.5. For the last three metrics, Csig reflects the signal distortion, Cbak estimates the background noise intrusiveness, and Covl predicts the overall quality of speech. They all produce values between 0 and 5. For all of these five metrics, a higher value stands for a better performance.

\subsection{Experimental setup}

The speech signals are transformed into spectrogram using Short-Time Fourier Transform (STFT). The length of frame and the hop-size are 512 points and 256 points, respectively. A 512-point Hanning window is used. In this way, an LPS with 257 frequency bins can be produced. The last frequency bin is eliminated so that the input and output of the network both have 256 frequency bins. Although this eliminating step is not necessary, it can help make a better comparison by keeping the same setup with the baseline methods such as UNet \cite{unet} and TCN. As for the output, the last frequency bin is padded with zero. Normalization is also applied,
\begin{equation}
  \tilde{Y}_{LPS} = (Y_{LPS} - U) / V
  \label{eq2}
\end{equation}
\begin{equation}
  \hat{S}_{LPS} = \tilde{S}_{LPS} * V + U
  \label{eq3}
\end{equation}
where $\tilde{Y}_{LPS}$ and $\tilde{S}_{LPS}$ are the input and output of the network, respectively. $U$ and $V$ is the mean value and standard deviation of all the noisy signals in the train data. The normalization and denormalization is performed on the frequency bin level, which means that $U$ and $V$ are vectors of 256 elements.

The loss function is the same with the one in UNet:
\begin{equation}
  loss = \frac{1}{T}\sum_{i=1}^{T}\sqrt{\frac{1}{F}\sum_{j=1}^{F}(S_{i,j} - \hat{S}_{i,j})^2}
  \label{eq4}
\end{equation}
where $S$ and $\hat{S}$ are clean and estimated LPS, respectively. $T$ is the number of frames and $F$ is the number of frequency bins.

\begin{table}[th]
  \caption{Parameters of TFCN}
  \label{tab:TFCN_parameters}
  \centering
  \begin{tabular}{ c c c c }
    \hline
    block & layer & channel number & kernel size \\
    \hline
    input & conv   & $1, 16$ & $5, 7$ \\
    \multirow{3}*{dilated} & conv\_0 & $16, 64$ & $1, 1$ \\
     & conv\_1 & $64, 64$ & $3, 3$ \\
     & conv\_2 & $64, 16$ & $1, 1$ \\
    output & conv & $16, 1$ & $1, 1$ \\
    \hline
  \end{tabular}

\end{table}

The hyper-parameters of TFCN are shown in Table~\ref{tab:TFCN_parameters}. The channel number is in shape (input channel, output channel), and the kernel size is in shape (height, width). Conv\_0, conv\_1 and conv\_2 stand for the three convolutional layers in one dilated block. There are $4$ repeated blocks in TFCN and each repeated block contains $8$ dilated blocks. The dilation rate of the $n^{th}$ dilated block within a repeated block is set to $2^n$, where $n$ starts from $0$.

The network is optimized with Adam optimizer. The speech signals are re-organized into segments with fixed length of 2s, i.e. 32000 sampling points, to form the batch data for training, while in the inference stage the intact utterances are fed into the network one by one. The initial learning rate is set to 0.001 and is halved if the validation loss is larger than the best loss for three continuous epochs. Early stop strategy is employed with patience of 10 epochs. The max epoch number is set to 100 while the training process usually stops within 50 epochs in practice.

\subsection{Results and discussion}

\begin{table}[th]
  \caption{Comparison with other methods}
  \label{tab:performances}
  \centering
  \begin{tabular}{ c c c c c c }
    \hline
    method & STOI & PESQ & Csig & Cbak & Covl \\
    \hline
    noisy    & 0.921 & 1.97 & 3.35 & 2.44 & 2.63 \\
    WaveNet \cite{wavenet}  & N/A & N/A & 3.62 & 3.23 & 2.98 \\
    C2F \cite{C2F}      & N/A & 2.73 & 3.94 & 3.35 & 3.33 \\
    UNet \cite{unet}    & 0.938 & 2.90 & 4.22 & 3.32 & 3.58 \\
    Affinity \cite{affinity} & N/A   & 3.04 & 4.30 & 3.42 & 3.69 \\
    \hline
    \hline
    TCN-time & 0.946 & 2.94 & 4.11 & \textbf{3.50} & 3.53 \\
    TCN-LPS  & 0.932 & 2.88 & 4.30 & 3.22 & 3.60 \\
    TFCN     & 0.943 & 3.00 & 4.38 & 3.44 & 3.71 \\
    TFCN-d   & \textbf{0.948} & \textbf{3.06} & \textbf{4.44} & 3.47 & \textbf{3.78} \\
    \hline
  \end{tabular}
\end{table}

\begin{table}[th]
  \caption{Number of parameters}
  \label{tab:param_num}
  \centering
  \begin{tabular}{ c c }
    \hline
    method & param number \\
    \hline
    UNet \cite{unet}    & 136.83M \\
    Affinity \cite{affinity} & 15.52M \\
    TCN-LPS  & 8.64M \\
    TFCN     & 0.09M \\
    TFCN-d   & 1.38M \\
    \hline
  \end{tabular}
\end{table}

The comparison with several state-of-the-art speech enhancement methods is shown in Table~\ref{tab:performances}. The baseline methods include a time-domain speech enhancement method based on WaveNet \cite{wavenet}, a convolutional neural network trained with coarse to fine optimization strategy \cite{C2F}, an LPS mapping method base on UNet \cite{unet} and its modified version \cite{affinity} which has two decoding branches and is optimized with subspace affinity loss. Besides, TCN-time is a time-domain speech enhancement method with multilayer encoder-decoder according to \cite{multilayerEncoder}. TCN-LPS directly use TCN described in Section 2.2 to estimate the clean LPS. TFCN is an implementation of the proposed structure with depth-wise convolution for the dilated convolutional layer. TFCN-d uses normal convolution instead and adds dense connections intra- and inter- repeated blocks to TFCN.

It can be seen that the proposed TFCN structures have better performances than the baseline methods. TFCN-d method achieves the best performance except for Cbak. TCN-time has the highest Cbak score, but get low scores in terms of Csig and Covl, which indicates that the time-domain method may have the best suppression capability of background noise but introduced more distortion to the speech signal. TFCN outperform TCN-LPS in terms of all the metrics, while the model size of TFCN is only about 1.1\% of TCN-LPS, as is shown in Table~\ref{tab:param_num}. This indicates that the proposed structure is more parameter-effective for LPS enhancement. Note that although the model size of TFCN-d is larger than TFCN, it is still significantly smaller than the other models mentioned above.

The reason that TFCN can achieve the excellent performance with only about 93k parameters lies in two aspects. One is the application of depth-wise separable convolutions, which is the same as TCN. The other one is that the 2-D convolutions are more efficient on modeling the LPS of speech signals, which allows the network to use fewer channels and further reduces the number of parameters.

The performance of the causal and semi-causal realization of TFCN-d is shown in Table~\ref{tab:causal}. It can be seen that the lack of the future information degrades the performance to a certain extent. However, the semi-causal realization with 304ms look ahead only suffers from a negligible degradation, which indicates that the proposed network can also be used in the situation that allows a reasonable latency.

\begin{table}[th]
  \caption{Performance of causal and semi-causal TFCN-d}
  \label{tab:causal}
  \centering
  \begin{tabular}{ c c c c c c }
    \hline
    method & STOI & PESQ & Csig & Cbak & Covl \\
    \hline
    Non-causal & 0.948 & 3.06 & 4.44 & 3.47 & 3.78 \\
    Semi (304ms) & 0.945 & 3.04 & 4.42 & 3.42 & 3.75 \\
    Semi (48ms)  & 0.942 & 2.98 & 4.38 & 3.39 & 3.70 \\
    Causal              & 0.940 & 2.91 & 4.32 & 3.36 & 3.63 \\
  \hline
  \end{tabular}
\end{table}

\section{Conclusions}

We propose TFCN for single channel speech enhancement. TFCN performs convolutional operations along both the time and frequency axis, and can effectively extract the temporal-frequential information. Experimental results show that TFCN can achieve an excellent performance with very few parameters. Using general convolutions instead of depth-wise convolutions and introducing the dense connections can further improve the performance. Experimental results also show that TFCN can work well both in causal and semi-causal situations. We will further improve the performance and parameter-efficiency of TFCN in our future work.

\bibliographystyle{IEEEtran}
\bibliography{mybib.bib}

\end{document}